\documentclass[letterpaper]{article} 
\usepackage{aaai25}  
\usepackage{times}  
\usepackage{helvet}  
\usepackage{courier}  
\usepackage[hyphens]{url}  
\usepackage{graphicx} 
\usepackage{natbib}  
\usepackage{caption} 
\usepackage{algorithm}
\usepackage{algorithmic}
\usepackage{booktabs}
\usepackage{bm}
\usepackage{amsmath}

\usepackage{newfloat}
\usepackage{listings}
\DeclareCaptionStyle{ruled}{labelfont=normalfont,labelsep=colon,strut=off} 
\floatstyle{ruled}
\newfloat{listing}{tb}{lst}{}
\floatname{listing}{Listing}
\title{Coverage Biases in High-Resolution Satellite Imagery}
\author{
    Vadim Musienko\textsuperscript{\rm 1},
    Axel Jacquet\textsuperscript{\rm 1},
    Ingmar Weber\textsuperscript{\rm 1},
    Till Koebe\textsuperscript{\rm 1}\thanks{Corresponding author: till.koebe@uni-saarland.de}
}
\affiliations{
    \textsuperscript{\rm 1}Saarland Informatics Campus, Saarland University, Saarbrücken, Germany\\
    vamu00001@stud.uni-saarland.de, axja00001@stud.uni-saarland.de, iweber@cs.uni-saarland.de, tkoebe@cs.uni-saarland.de
}

\usepackage{bibentry}

\begin{document}

\maketitle

\begin{abstract}\footnote{To appear in the proceedings of the ICWSM'25 Workshop on Data for the Wellbeing of Most Vulnerable (DWMV). Please cite accordingly.}
Satellite imagery is increasingly used to complement traditional data collection approaches such as surveys and censuses across scientific disciplines. However, we ask: Do all places on earth benefit equally from this new wealth of information? In this study, we investigate coverage bias of major satellite constellations that provide optical satellite imagery with a ground sampling distance below 10 meters, evaluating both the future on-demand tasking opportunities as well as the availability of historic images across the globe. 
Specifically, forward-looking, we estimate how often different places are revisited during a window of 30 days based on the satellites' orbital paths, thus investigating potential coverage biases caused by physical factors. We find that locations farther away from the equator are generally revisited more frequently by the constellations under study. Backward-looking, we show that historic satellite image availability -- based on metadata collected from major satellite imagery providers -- is influenced by socio-economic factors on the ground: less developed, less populated places have less satellite images available, as demonstrated by a Gini coefficient of $0.64$. Furthermore, in three small case studies on recent conflict regions in this world, namely Gaza, Sudan and Ukraine, we show that also geopolitical events play an important role in satellite image availability, hinting at underlying business model decisions. These insights lay bare that the digital dividend yielded by satellite imagery is not equally distributed across our planet.
\end{abstract}

\section{Introduction}\label{section-intro}

In recent years, advances in satellite technology and data processing capabilities have helped remote sensing products such as satellite imagery revolutionize the way we observe and understand the Earth's surface, e.g. by augmenting traditional data collection methods such as surveys and censuses across a wide range of scientific disciplines. Applications range from environmental monitoring and disaster response to urban planning and agricultural management. The ability to capture and process vast amounts of data remotely has made satellite imagery a cornerstone of contemporary decision-making in both research and policy contexts. However, the global reach of satellite technology raises an important question: Does the availability of satellite imagery truly benefit all regions of the world equally? While satellites orbiting the Earth may appear to provide unbiased data collection opportunities, the reality is far more complex: Differences in business models of satellite operators and geophysical trade-offs of orbital paths influence both possible revisitation rates as well as the availability of historic imagery. These disparities have significant implications for how satellite data can be leveraged, potentially exacerbating existing inequalities in information access and utilization. We summarize the research questions (RQs) that motivate this paper as follows:

\begin{enumerate}
    \item \textbf{RQ1:} Are satellite images equally available for all regions in the world or do coverage biases exist?
    \item \textbf{RQ2:} If those biases exist, what are the determinants of satellite image availability?
\end{enumerate}

Disparities in satellite image availability have already been explored in previous works. For example, \citet{lesiv2018characterizing} show that the very high resolution (VHR) satellite imagery on Google Earth and Bing Maps is unevenly global distributed throughout the world, with better coverage in the USA, Europe and India. This bias is further reflected in open-source image datasets, which, according to \citet{shankar2017no} exhibit amerocentric and eurocentric representation, potentially impacting classifier performance in developing regions. Similarly, \citet{kumar2018google} finds that processing platforms that help end-user to large amounts of imagery such as Google Earth Engine (GEE) are mainly used by institutions in developed nations. While initiatives such as the DeepGlobe 2018 Satellite Image Understanding Challenge aim to overcome entry barriers, inequities still remain \citep{demir2018deepglobe}. Although recent advances driving down GSD with ever-higher resolutions of images, the most used satellite constellations remain Landsat and Sentinel \citep{kumar2018google}.  Here, the evolution of data sharing policies for environmental satellite data has seen a shift towards open access, potentially benefiting various environmental and social applications (cf. \citet{wulder2012opening, turner2015free, olbrich2018open}). \citet{sudmanns2020assessing} assess the global spatio-temporal coverage and availability of Sentinel-2 satellite imagery and find that higher revisitation rates do not necessarily lead to more cloud-free imagery available -- an essential precondition for optical satellite imagery to be of further use. In contrast, \citet{li2017global} find that combining data from multiple satellites, such as Landsat-8 and Sentinel-2A/B, can significantly improve temporal revisit intervals and observation frequency, thus enhancing opportunities for cloud-free surface observations. 

In general, applications for satellite imagery are plentiful: Among others, as described by \citet{voigt2016global} and \citet{rufener2024estimation}, rapid emergency mapping is extensively used to assess major disaster situations, with Asia and Europe being the most intensively monitored regions. To improve global land cover maps, initiatives like OpenEarthMap -- a benchmark dataset for high-resolution land cover mapping, which consists of 2.2 million segments covering 97 regions across 6 continents -- have emerged \citep{xia2023openearthmap}. Satellite imagery is also used to refine poverty mappings and demographic projections (cf. \citet{koebe2022intercensal, pokhriyal2017combining}). These satellite-based earth observations have potentially significant impacts on society and policy, as discussed in \citet{onoda2017satellite}.

This study seeks to go beyond Google Maps and Bing Maps as done by \citet{lesiv2018characterizing} to highlight inequitable access to information by investigating the spatial and temporal coverage biases inherent in major satellite constellations providing optical satellite imagery at resolutions with a ground sampling distance (GSD) below 10 meters, namely Planetscope, Maxar, One Atlas, Imagesat, 21at and Capella Space. While we recognize that both the Landsat as well as the Sentinel programme are essential sources of information in the remote sensing community, we do not consider them in this analysis for mainly two reasons: First, with resolutions of 10m and above, both Landsat and Sentinel support use cases that are distinct to those of high-resolution imagery providers. Second, as both Landsat and Sentinel are government-led programmes, they follow a systematic (''gotta catch'em all'') capture model, irrespective of specific user demands. Thus, although we expect that physical factors such as orbital paths still influence forward-looking coverage for those two programmes, we expect that the backward-looking coverage should not be influenced by socio-economic factors.

We break our analysis down into two dimensions: forward-looking and backward-looking. Forward-looking, we investigate whether revisitation rates structurally differ across different places on earth. Here, the revisitation rate provides an upper bound of potential tasking opportunities for a given area in a specific time interval. Backward-looking, we examine metadata from major satellite imagery archives, namely Up42, Maxar and Planet to identify patterns and coverage biases in historic image availability and link it to potential determinants such as population density and income, but also the geo-location on the subnational level. Together, these analyses provide a comprehensive assessment of how the "digital dividend" afforded by satellite imagery is distributed across the globe. Thus, our contributions in this study are twofold:

\begin{enumerate}
    \item \textbf{C1:} We compare actual image availability (backward-looking) to its theoretical upper bound (forward-looking), thereby showcasing the economic reasoning behind data availability.
    \item \textbf{C2:} We investigate socio-economic and geo-physical determinants of historic satellite image availability at the subnational level on a global scale.
\end{enumerate}

Together, these analyses provide a comprehensive assessment of how the "digital dividend" afforded by satellite imagery is distributed across the globe. We hope this study informs downstream analyses that quantify the economic, political and social impact of these biases and ultimately guide data sharing and development policies towards enabling equitable access globally.

\section{Data}\label{section-data}

Satellites are mostly organized into constellations, which are groups of satellites working together for a set of specific purposes. The most well-known example is the GPS satellites used for positional location. Therefore, in order to estimate satellite coverage of certain locations, it is essential to examine entire constellations rather than individual satellites. We consider the following constellations in our analysis: Planet Labs operates three satellite constellations to provide imagery of Earth: Dove (GSD 3-5m), RapidEye (retired, GSD 5m) and Skysat (GSD 50cm)\footnote{\url{www.planet.com/our-constellations/}}. Maxar runs two constellations: WorldView (consisting of the satellites WorldView-1, WorldView-2, WorldView-3, GeoEye-1 with GSDs between 30-50cm) and WorldView Legion (GSD 30cm)\footnote{\url{https://www.maxar.com/maxar-intelligence/constellation}}. 21st Century Aerospace Technology (21AT) runs the TripleSat constellation (GSD 80cm)\footnote{\url{https://www.21at.sg/productsservices/triplesat-constellation/}}. The EROS constellation (GSD 70-180cm) is operated by 
ImageSat International\footnote{\url{https://imagesatintl.com/}}. Airbus groups its Pléiades, Pléiades Neo, SPOT and Vision-1 and SAR satellites into one constellation (optical satellites have a GSD 30-150cm)\footnote{\url{https://space-solutions.airbus.com/imagery/our-optical-and-radar-satellite-imagery/}}.

In this paper, we examine four main underlying data sources: First, Two-Line Elements (TLEs) from \url{space-track.org} to determine theoretical satellite coverage in the forward-looking part of the paper; Second, Spatiotemporal Asset Catalogs (STACs) of major satellite operators and redistributors of historic satellite imagery to assess actual image availability in the backward-looking part of the paper; Third, the global dataset of \citet{ton2024global} on socio-economic characteristics at the subnational level to investigate the socio-economic and geophysical determinants of potential coverage biases; And fourth, the Subnational Human Development Index (SHDI) published by Global Data Lab \citep{smits2019subnational} to complement the global dataset by \citet{ton2024global}.

TLE sets are a compact data format describing orbital parameters of satellites that are widely used for orbit prediction and tracking. Each TLE set contains the necessary information to describe the orbit of a satellite at a specific moment in time and is structured into two lines of data, hence the name. The first line provides basic information, e.g. timestamp, identifier and basic motion parameters. The second line provides orbital elements such as the shape of the orbit (\textit{eccentricity}) and the orbital tilt relative to the equator (\textit{inclination}). These parameters are typically derived from observations and maintained by organizations such as United States Space Forces and partially provided to the public via websites such as space-track.org.

STACs are an open standard for organizing and sharing geospatial data and widely used in geospatial data systems to improve accessibility and interoperability of large-scale datasets. STAC organizes geospatial data as items in form of hierarchically structured catalogs that share standardized metadata fields, which, however, can be further extended with custom fields. We retrieve metadata on available imagery from Maxar's STAC \footnote{\url{https://developers.maxar.com/docs/discovery/api/query-the-catalog}} directly and for the other satellite operators via the STAC of Up42\footnote{\url{https://docs.up42.com/developers/api-stac}} as a major redistributor of satellite imagery. For Planet imagery, we retrieve only metadata statistics from their STAC API\footnote{\url{https://developers.planet.com/docs/apis/data/reference/}} in order to keep the amount of data manageable.

\citet{ton2024global} present GLOPOP-S, a synthetic global population dataset on 7.3 billion individuals and 2 billion households. This open-source dataset includes socio-economic attributes such as age, gender, income, wealth, education, and settlement type by integrating data from the Luxembourg Income Study (LIS) and the Demographic and Health Survey (DHS) program. Data and code are freely available for download\footnote{\url{https://dataverse.harvard.edu/dataset.xhtml?persistentId=doi:10.7910/DVN/KJC3RH}}. 

The subnational Human Development Index (SHDI), curated by Global Data Lab (an independent research center at Radboud University), is a composite index that is based on the three dimensions 'education', 'health' and 'standard of living'. Underlying indicators include \textit{years of schooling}, \textit{life expectancy at birth} and \textit{gross national per capita income}. For further methodological details, we refer to \citet{smits2019subnational} and to the website of the Global Data Lab\footnote{\url{https://globaldatalab.org/shdi/about/}}.

To complement the metadata-based analysis of satellite imagery availability, we also account for atmospheric conditions that affect image usability. Specifically, we retrieve hourly cloud cover data for each GDL region using the Open-Meteo Archive API \citep{Zippenfenig_Open-Meteo}, based on a region's centroid and aggregated over the year 2023.

\section{Methodology}\label{section-methods}

In this study, we investigate coverage biases in historic satellite image availability (backward-looking) and in potential future tasking opportunities (forward-looking). 

\subsection{Orbital path estimation}\label{sec:meth_orbital}

\begin{figure}[t!]
    \centerline{\includegraphics[width=1.00\columnwidth]{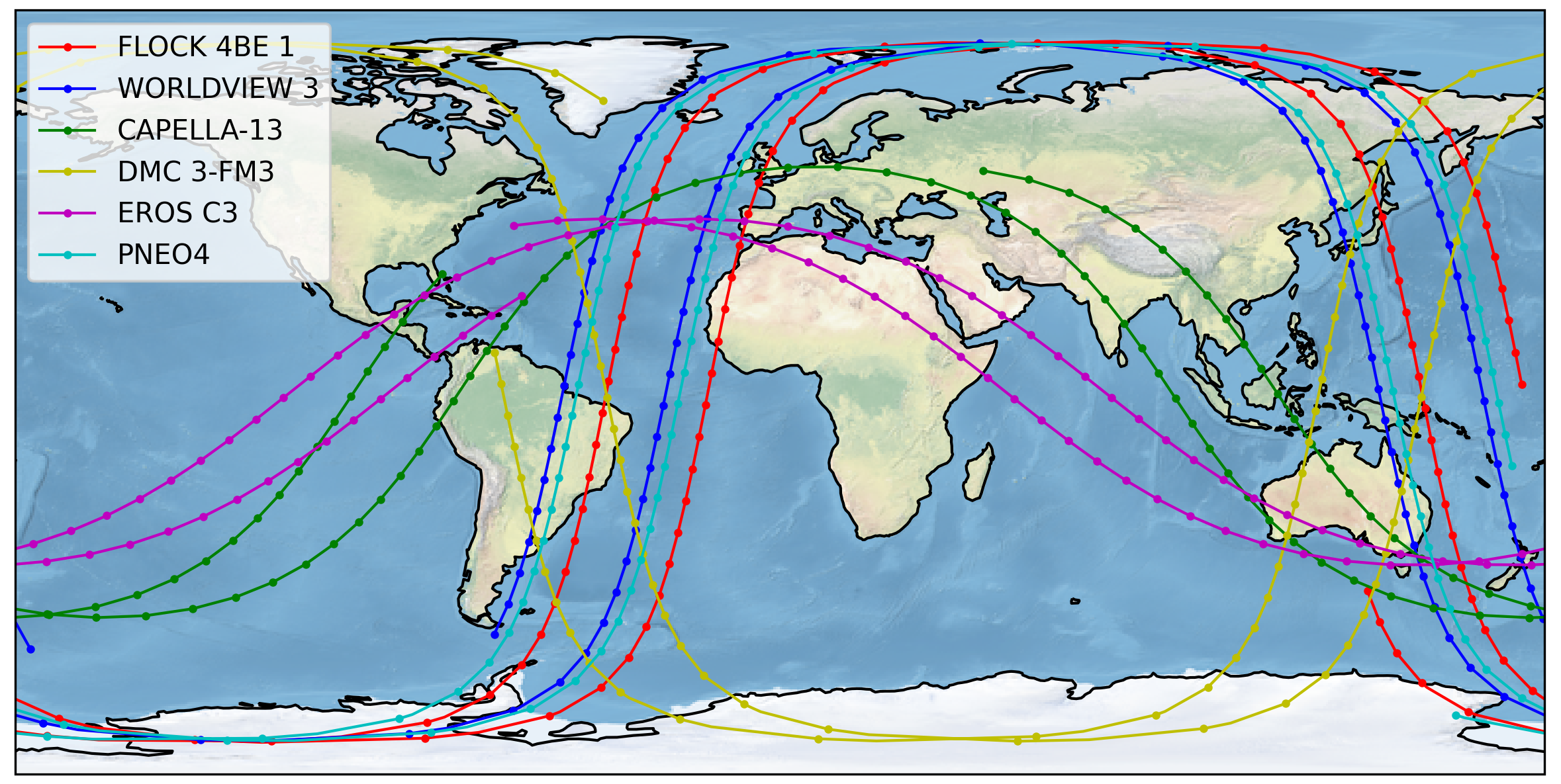}}
    \caption{Tracks from satellites of different constellations during a three-hour time window.}
    \label{fig:orbits_different_constellations}
\end{figure}

In the forward-looking case, we use TLE data of major satellite constellations to estimate revisitation rates by simulating each satellite’s position over time. Figure \ref{fig:orbits_different_constellations} illustrates the orbital paths of an individual satellite from each studied constellation for a three-hour time window (11:10:15 to 14:10:15 UTC) on 29th January 2024.

For path estimation, we use the Skyfield Python library \citep{2019ascl.soft07024R} to propagate satellite positions from publicly available Two-Line Element (TLE) data. We map these continuous orbital paths at intervals of 1 minute to both subnational administrative areas and a gridded earth surface with an edge length of approx. 500 kilometers. We approximate the width of potential scenes captured by a satellite with a 250 kilometer buffer around its predicted orbital path. Finally, we count the times the buffered paths include the centroid of a given tile within the time window of interest and interpret the resulting count as the theoretic revisitation rate at that tile.

\subsection{Regression analysis}\label{sec:meth_linear}

Backward-looking, we extract metadata from the image archives of the five constellations studied from Up42, Maxar and Planet for the years 2017 to 2023. The metadata include information on the time of image capture, on the GSD, and geographic extent of the captured image. We assign images both to sub-national boundaries (admin-level 1) as well as to an earth grid by the centroid of the geographic extent. While this approach ignores the fact that some satellites partially cover multiple regions in one capture, it is computationally efficient and we assume the effect of this simplification to diminish given the large time window of study. To further explore spatial patterns in the historic image availability, we categorize satellite images along the GSD, namely in following ranges: 0–0.5 meters, 0.5–1.0 meters, 1.0–3.0 meters, and above 3.0 meters. Furthermore, we hypothesize about potential factors that influence historic satellite image availability. As most socio-economic data is available along administrative boundaries, we consider admin-level 1 regions as our geographies of interest for bias investigation. Besides the area size of a given region (in sqkm), we hypothesize that more populated regions may be more prominent in image archives compared to sparsely populated ones. We further hypothesize that similar holds for economically more viable regions. As a control for the theoretical bias due to a satellite's orbital path, we add the longitude and the latitude of a region's centroid to the regression. Also, as the usefulness of optical satellite imagery heavily relies on the absence of clouds, we consider a region's average cloud coverage as potentially relevant indicator of historic image availability. A full list of variables considered in the regression can be found in the Appendix \ref{appendix:variables}. To identify significant factors that influence the sub-national image count, we formulate it as a ordinary least squares (OLS) linear regression problem, generically defined as

\begin{equation}
    y_i = \alpha + \bm{X_i\beta} + \epsilon_i,
\end{equation}

\noindent where $i \in \{1, \dots, N\}$, $y_i$ represents the count of available historic images for region $i$, $\alpha$ the intercept, $\bm{\beta}$ a $n \times 1$ vector of coefficients, $\bm{X_i}$ is a  $1 \times n$ vector of region-level covariates and $\epsilon_i$ the region-specific residual term.

In the last step of the backward-looking analysis, we compare the results of the regression with the Gini coefficient -- a measure of inequality that, in our setup, indicates how equally distributed historic satellite images are across the different regions in the world, sorted by their respective Subnational HDI.

In three descriptive case studies, we further investigate if geopolitical events influence (historic) satellite image availability by visualizing image counts around the onset of major recent violent conflicts, namely the Russian invasion in Ukraine from February 2022, the civil war in Sudan starting April 2023 and the Israel-Gaza war from October 2023 onwards.

\section{Results}\label{section-results}

\subsection{Forward-looking}\label{section-forward}

\begin{figure}[ht!]
    \centering
    \includegraphics[width=\linewidth]{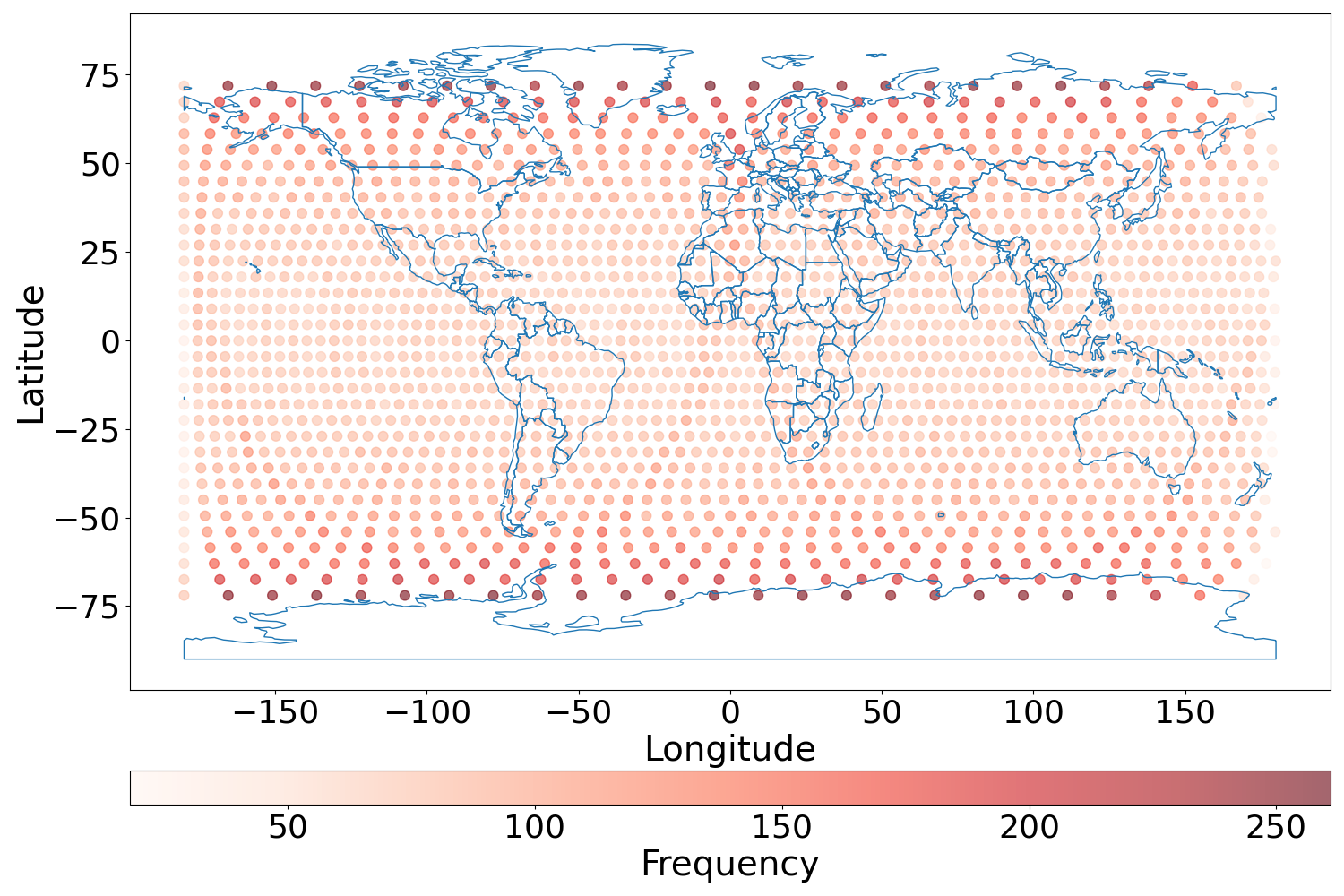}
    \caption{Skysat revisitation density within 30 days based on SkySat satellite C19.}
    \label{fig:revisitation-density}
\end{figure}

Figure \ref{fig:revisitation-density} shows the revisitation rate at each tile of a gridded representation of earth using the example of the satellite \textit{C19} from the Skysat constellation.

\begin{figure}[ht!]
    \centering
    \includegraphics[width=1.1\linewidth]{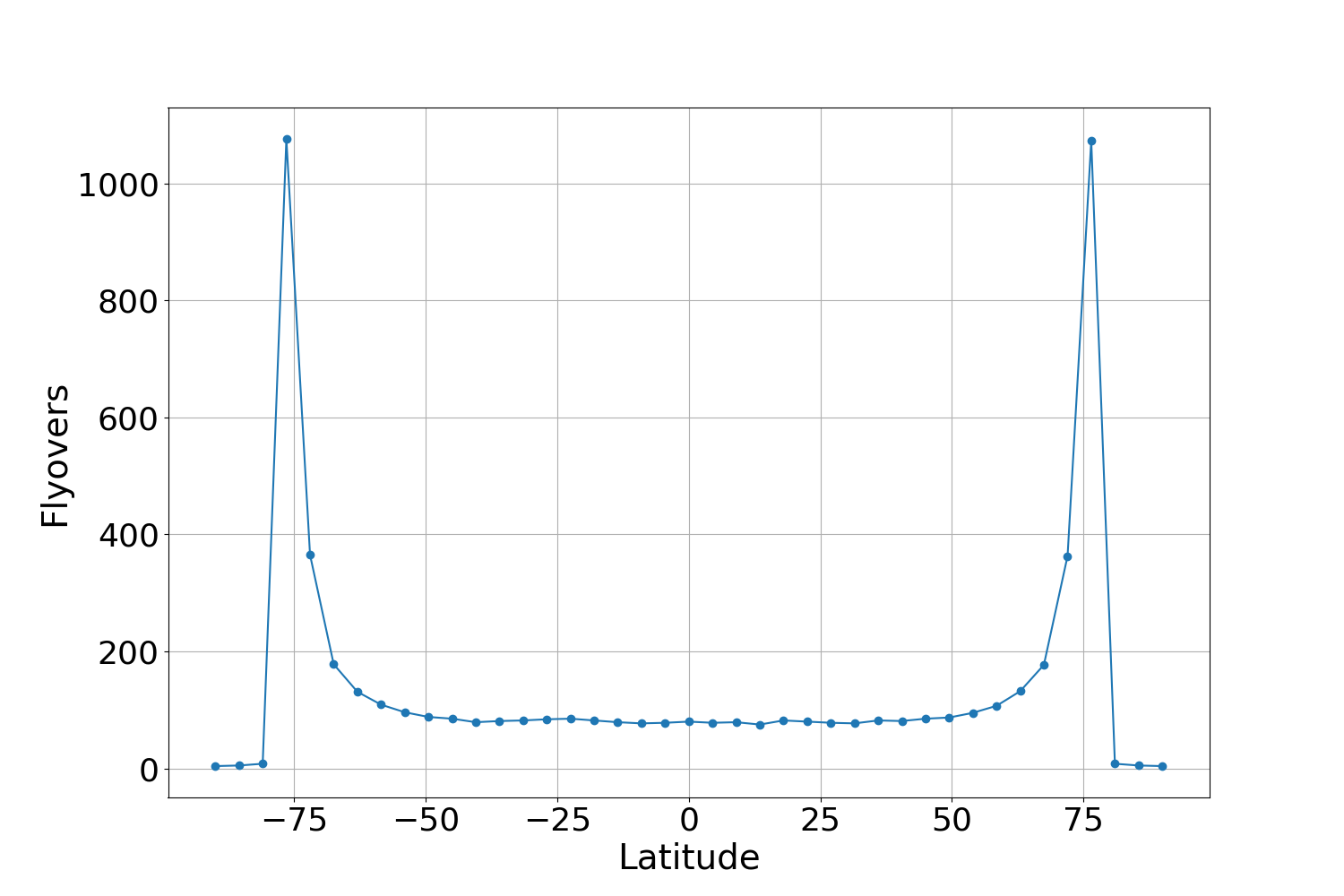}
    \caption{Latitude - Revisitation Density - Plot based on SkySat satellite C19.}
    \label{fig:lat-revisit}
\end{figure}

In this figure, a steady decline of the revisitation rate along the latitude towards the equator is evident: Around the longitudes, the revisitations are uniformly spread out. Around the latitudes, the revisitations peak in a ring around the poles with about 1000 revisits per 30 days (not displayed to keep the scale legible). Near the equator, there are significantly fewer revisits by the SkySat constellation, with about 80 revisits per 30 days. However, as showcased in Figure \ref{fig:lat-revisit}, the revisitation rate remains fairly constant between a latitude of -50° to 50°. As most observed constellations follow a sun-synchronous orbit, we assume the revisitation density of the depicted Skysat satellite to be by and large representative for most satellites in the sample. In Table \ref{tab:monthly_revisits}, we investigate whether the revisits per continent differ across resolutions the respective satellites capture.

\begin{table}[ht]
\centering
\small
\caption{Monthly revisits by continent and resolution (30-day estimate).}
\label{tab:monthly_revisits}
\begin{tabular}{l@{\hspace{6pt}}r@{\hspace{6pt}}r@{\hspace{6pt}}r@{\hspace{6pt}}r}
\toprule
\textbf{Continent} & \textbf{0.3–0.5m} & \textbf{0.5–1.0m} & \textbf{1.0–3.0m} & \textbf{$>$3.0m} \\
\midrule
Africa       & 12,660  & 55,080  & 72,540   & 1,866,060  \\
America      & 28,620  & 117,330 & 161,310  & 4,097,040  \\
Asia/Pacific & 19,290  & 90,660  & 107,940  & 2,725,440  \\
Europe       & 19,950  & 77,790  & 113,160  & 2,814,930  \\
\bottomrule
\end{tabular}
\end{table}

The table shows that the vast majority of monthly revisits occur at coarse resolutions ($>3.0$ m), with America and Europe contributing the highest revisit counts overall. Here, it needs to be noted that in our analysis, Greenland belongs to the American continent and all of Russia is assigned to Europe. 



We observe that Africa shows higher revisit rates per $\text{km}^2$ across all GSD ranges than the other continents. We expect two mechanisms at play here: On one hand, Africa is mostly centered around the equator, thus reducing the revisitation density (cf. Figure \ref{fig:lat-revisit}). On the other hand, the African landmass is compact and rather spread out across the latitudes than longitudes, aligning more with the orbital paths of satellites, thus increasing the number of potential captures. While the latter effect may dominate the former on a continental scale, this does not apply on the subnational/local-level.

\subsection{Backward-looking}\label{section-backward}

\begin{figure}[t]
    \centering
    \includegraphics[width=\linewidth]{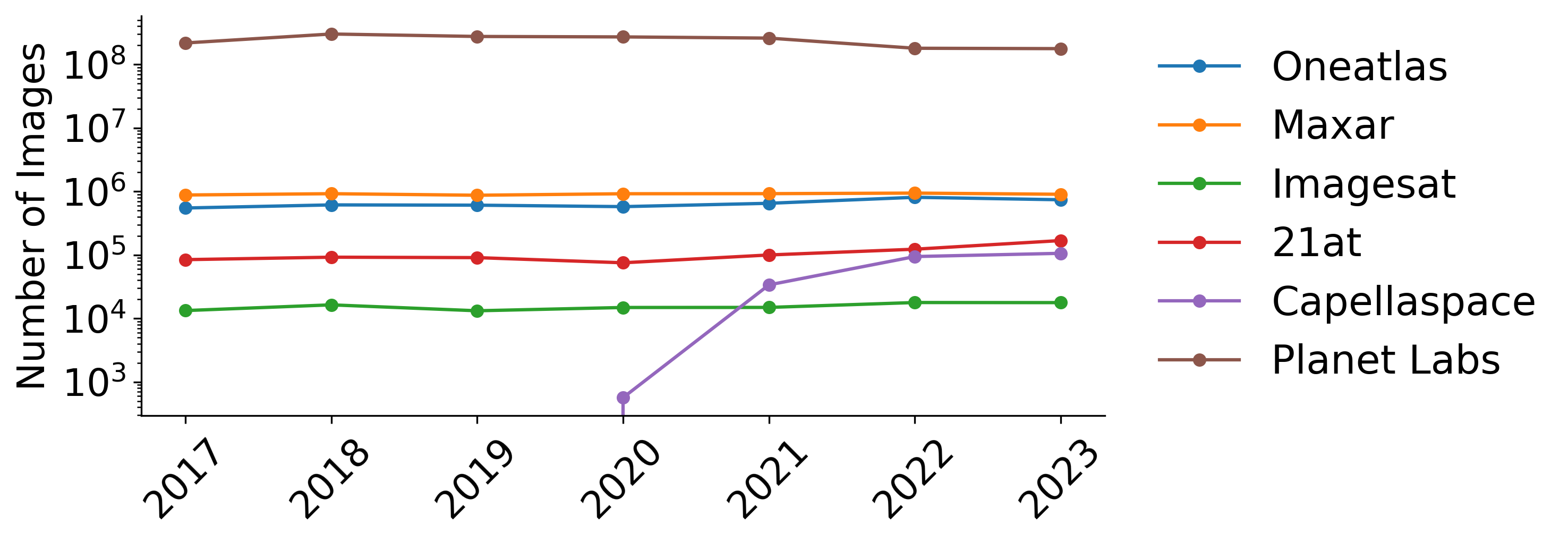}
    \caption{Counts of available historic satellite imagery for the years 2017--2023 by provider.}
    \label{fig:line_graph_number_of_satellite_images_over_time_by_provider}
\end{figure}

\begin{figure}[t]
    \centering
    \includegraphics[width=\linewidth]{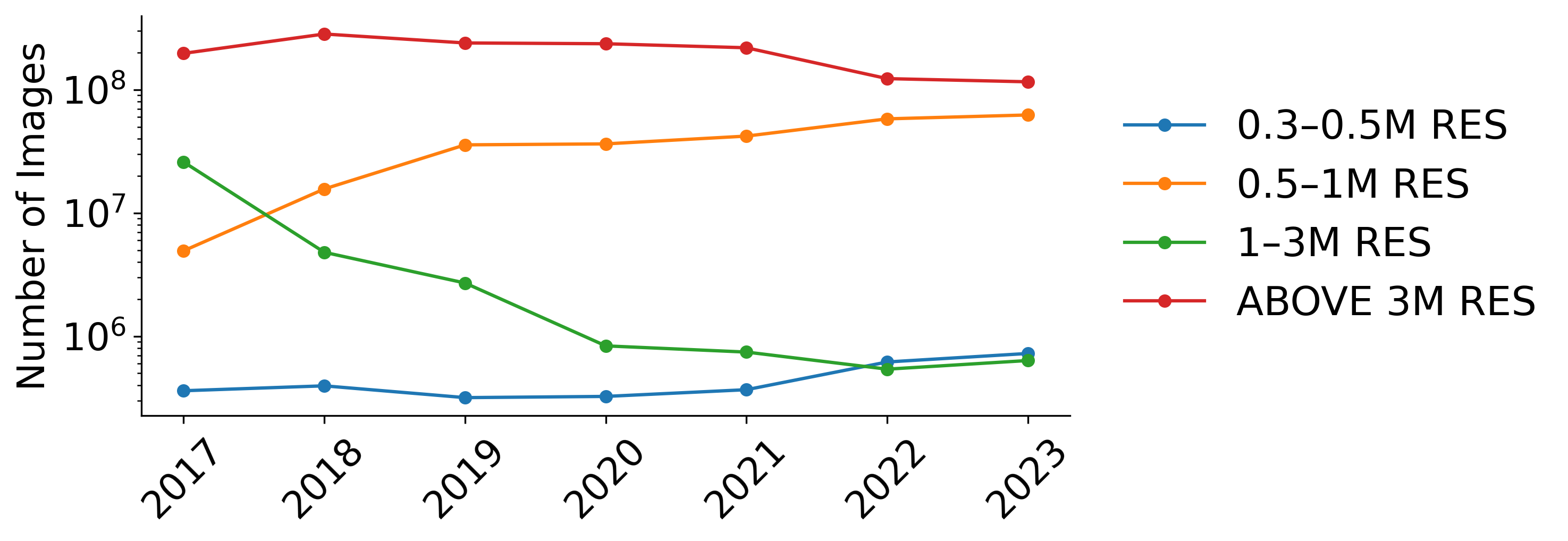}
    \caption{Counts of available historic satellite imagery for the years 2017--2023 by resolution.}
    \label{fig:line_graph_number_of_satellite_images_over_time_by_resolution}
\end{figure}

\begin{figure}[t]
    \centering
    \includegraphics[width=\linewidth]{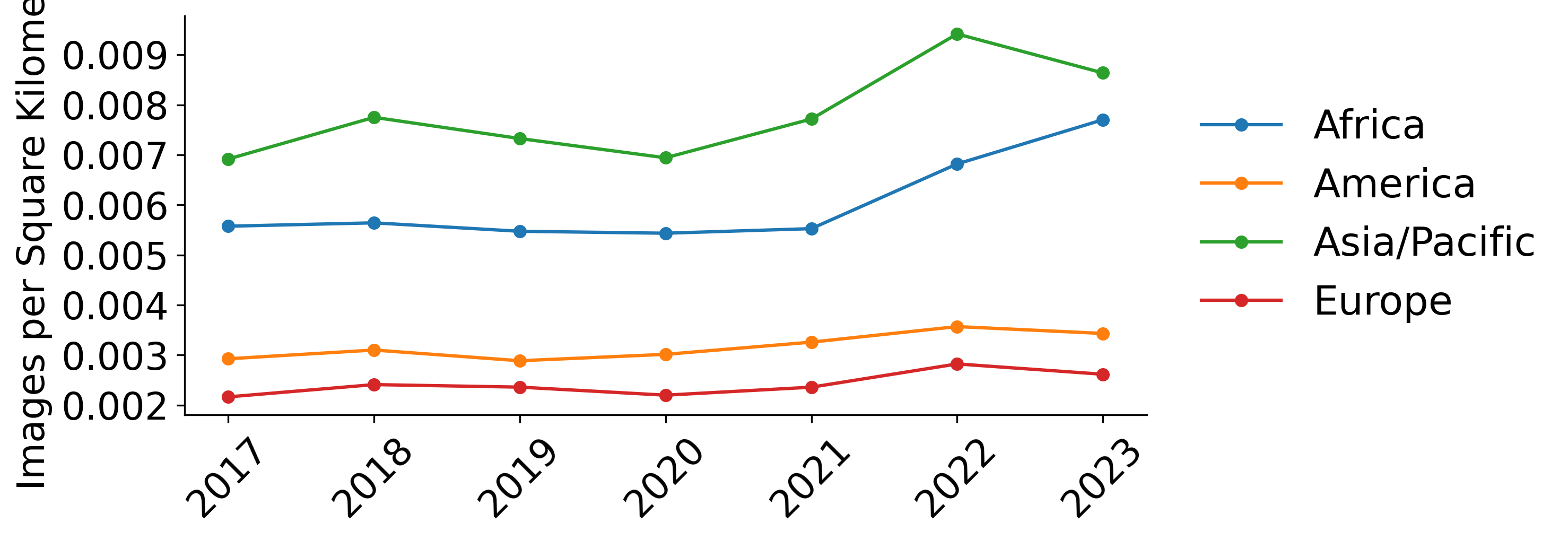}
    \caption{Counts of available historic satellite imagery for the years 2017--2023 by continent.}
    \label{fig:image_counts_by_year_and_continent}
\end{figure}

In a next step, we compare the revisit rates (as described in the forward-looking case), which we consider to represent the potential image availability, with the actual image availability for the previous years (as described in Section \ref{sec:meth_linear}). Figures \ref{fig:line_graph_number_of_satellite_images_over_time_by_provider}--\ref{fig:image_counts_by_year_and_continent} shows the historic image availability across years, disaggregated by provider, resolution and continent, respectively.

\begin{figure*}[ht]
    \centering
    \includegraphics[width=\textwidth]{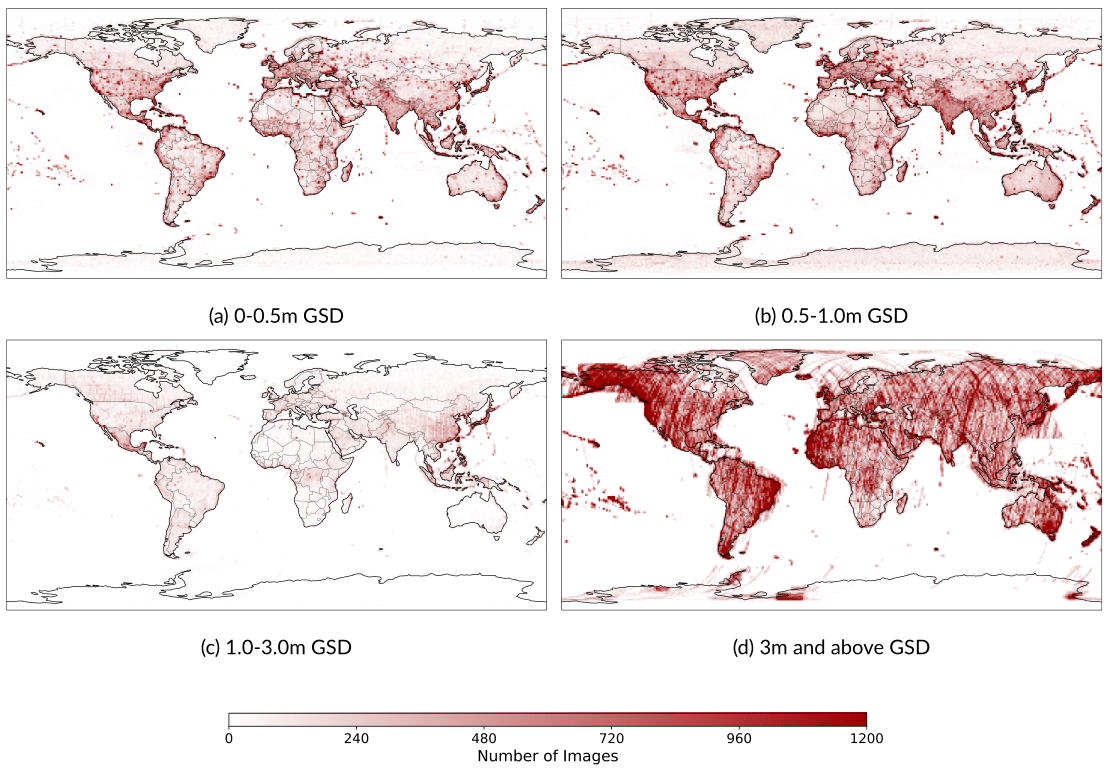}
    \caption{Number of publicly available historic satellite images from the years 2017--2023, by resolution range (right-open).}
    \label{fig:resolution_heatmap_grid}
\end{figure*}

We observe that Planet by far hosts the most historic satellite images. This is also reflected in the image availability at a resolution of 3m and above as the main Planet constellation consisting of Dove satellites capture images with a GSD of 3-5m. Also, the trend towards very-high resolution (VHR) images, i.e. GSD < 1m, is clearly visible with the below 1m resolution images seeing an increase in availability over time, while above 1m resolution images decrease simultaneously. While technological advances provide the necessary conditions for VHR imagery, another major contributing factor is an increase in demand for VHR tasking as tasked images are in many cases resold as archived images thereafter at a fraction of tasking prices. We assume that this business model leads to a biased availability of historic satellite imagery, as satellite tasking still remains a costly endeavor. In Figure \ref{fig:resolution_heatmap_grid}, we map the centroids of the archived images to the gridded earth to investigate their fine-granular spatial distribution.

In this figure, we see that VHR imagery availability appears to be higher in populated places such as India and Europe than in less populated places such as the Sahara, Siberia, or the Arctic and Antarctic regions. Interestingly, historic satellite imagery is also available for certain regions of the international water, mainly in the Pacific ocean in the area around French Polynesia. Comparing the average monthly revisit rate with the average monthly historic image counts on the level of continents (see Tables \ref{tab:revisits_v_historic_africa}--\ref{tab:revisits_v_historic_europe}), this link still persists.

\begin{table*}[tb]
\centering
\small
\caption{Number of daily revisits and average daily historic images, by continent and resolution.}
\label{tab:detailed_table}
\begin{tabular}{llccc}
\toprule
\textbf{Continent} & \textbf{Resolution (m)} & \textbf{Avg. Daily Revisits} & \textbf{Avg. Daily Historic Images} & \textbf{Avg. Ratio} \\
\midrule
Africa & 0.3–0.5 & 422 & 64 & 0.141 \\
Africa & 0.5–1.0 & 1836 & 180 & 0.094 \\
Africa & 1.0–3.0 & 2418 & 67 & 0.028 \\
Africa & \(>3.0\) & 62202 & 564 & 0.010 \\
\midrule
America & 0.3–0.5 & 954 & 116 & 0.116 \\
America & 0.5–1.0 & 3911 & 469 & 0.109 \\
America & 1.0–3.0 & 5377 & 196 & 0.037 \\
America & \(>3.0\) & 136568 & 1578 & 0.012 \\
\midrule
Asia/Pacific & 0.3–0.5 & 643 & 135 & 0.198 \\
Asia/Pacific & 0.5–1.0 & 3022 & 423 & 0.129 \\
Asia/Pacific & 1.0–3.0 & 3598 & 210 & 0.059 \\
Asia/Pacific & \(>3.0\) & 90848 & 978 & 0.010 \\
\midrule
Europe & 0.3–0.5 & 665 & 70 & 0.109 \\
Europe & 0.5–1.0 & 2593 & 187 & 0.086 \\
Europe & 1.0–3.0 & 3772 & 86 & 0.023 \\
Europe & \(>3.0\) & 93831 & 934 & 0.010 \\
\bottomrule
\end{tabular}
\end{table*}

Africa and Asia-Pacific, by far the two most populated continents, show the highest rates of very high-resolution (i.e. $0.3-0.5\text{m}$) images to potential revisits. However, overall differences in the ratios by resolution across continents are low. Furthermore, this pattern is less evident for mid-resolution (i.e. $0.5-3\text{m}$) imagery and almost non-evident for lower-resolution (i.e. $>3\text{m}$) imagery. The latter is might be explained by the fact that the Planetscope constellation follows a capturing strategy similar to the government-led "gotta catch 'em all" approach of the Landsat and Sentinel programmes. In order to identify drivers of historic satellite imagery availability, we combine average monthly subnational image counts with additional, globally available information, e.g. related to regional weather and socio-economic conditions (see Table~\ref{tab:regression_results_main}).

\begin{table*}[t]
\centering
\small
\begin{tabular}{lcccc}
\toprule
& \multicolumn{4}{c}{\textit{Dependent variable: Historic image count}} \\
\cmidrule(lr){2-5}
& (1) & (2) & (3) & (4) \\
\midrule
Constant & -0.013$^{**}$ & -0.005$^{}$ & -0.010$^{***}$ & -0.013$^{***}$ \\
& (0.005) & (0.006) & (0.002) & (0.002) \\
Subnational HDI & 0.045$^{***}$ & -0.017$^{}$ & 0.011$^{***}$ & 0.007$^{**}$ \\
& (0.007) & (0.011) & (0.003) & (0.003) \\
Latitude (abs.) & & 0.053$^{***}$ & 0.006$^{***}$ & 0.010$^{***}$ \\
& & (0.007) & (0.002) & (0.002) \\
Longitude (abs.) & & 0.056$^{***}$ & 0.014$^{***}$ & 0.013$^{***}$ \\
& & (0.005) & (0.002) & (0.002) \\
\# of households & & & 0.075$^{***}$ & 0.076$^{***}$ \\
& & & (0.005) & (0.005) \\
Area size & & & 1.035$^{***}$ & 1.035$^{***}$ \\
& & & (0.008) & (0.008) \\
Cloud coverage & & & & 0.008$^{***}$ \\
& & & & (0.002) \\
\midrule
Observations & 1726 & 1726 & 1726 & 1726 \\
$R^2$ & 0.021 & 0.086 & 0.913 & 0.914 \\
Adjusted $R^2$ & 0.021 & 0.084 & 0.913 & 0.914 \\
Residual Std. Error & 0.050 (df=1724) & 0.049 (df=1722) & 0.015 (df=1720) & 0.015 (df=1719) \\
F Statistic & 37.556$^{***}$ (df=1; 1724) & 53.919$^{***}$ (df=3; 1722) & 3625.426$^{***}$ (df=5; 1720) & 3052.306$^{***}$ (df=6; 1719) \\
\bottomrule
\textit{Note:} & \multicolumn{4}{r}{$^{*}$p $<$ 0.1; $^{**}$p $<$ 0.05; $^{***}$p $<$ 0.01} \\
\end{tabular}
\caption{OLS: Regressing the number of publicly available historic satellite images (without planet images) from the years 2017--23 on socio-economic and physical factors. All variables are normalized.}
\label{tab:regression_results_main}
\end{table*}

Naturally, our control variable \textit{Area size} is the major driver of historic image counts across the 1726 subnational regions under study. By controlling for the area size, the number of households in a region rather reflect the population density in a given region, which appears to have a positive and significant influence on the image count. In other words, rural populations are disadvantaged when it comes to the access to satellite imagery with a resolution below $10\text{m}$. We also observe a small, but positive and significant effect of the subnational HDI on the normalized number of historical images available on a subnational level. The effect is robust across three of four model specifications. In other words, less developed regions have slightly less historic images available, keeping other things equal. However, the normalized subnational HDI explains only about 2\% of the variation observed in the normalized historic image counts. Thus, the overall impact of socio-economic conditions are quite limited.

From the forward-looking perspective, one would expect that the absolute latitude value of a region positively influences image availability (i.e. the closer to the equator, the less images). Our regression results confirm this hypothesis. Interestingly, there appears to be also a significant positive effect of the longitude of a region. We assume this significance is caused by some omitted variables since, for example, socio-political divides, expressed through terms such as the 'Western Hemisphere' or the 'Western World' may loosely be represented along the longitude, but likely correlate with important factors for image availability such as a region's economic power. Counterintuitively, cloud coverage appears to positively and significantly affect image availability. One potential explanation is that local economic hubs are often located along coastal lines, which are more prone to cloud cover compared to inland areas, however, further investigations into the potential underlying causal mechanisms are needed.

To explore the robustness of the observed relationships, we conducted multiple ablation studies and present the results in the Appendix. In our main regression, we exclude imagery from Planet's dove constellation for two reasons: since Planet is also hosting by far the most historic images due to its "gotta catch'em all" strategy (cf. Figure \ref{fig:line_graph_number_of_satellite_images_over_time_by_provider}), we expect the determinants of historic image availability for Planet to be structurally different from other providers and also to dominate the reported relationships. This is supported by the results for all available historic images (incl. Planet) and for Planet-only images are provided in Table \ref{tab:total_regression_results_appendix_all} and Table \ref{tab:total_regression_results_appendix_planet}, respectively, in the Appendix. Surprisingly, we observe that by including historic Planet images, the effect of the subnational HDI and the number of households becomes insignificant and/or negative, while the effects of the remaining variables remain largely unchanged. Assuming that Planet's dove constellation captures virtually all possible landmass images, one would expect the coefficients of the variables \textit{Cloud coverage}, \textit{\# of households} and \textit{subnational HDI} to become insignificantly different from zero, especially when looking at the Planet-only results. Possible reasons for this are multiple: First, omitted variable bias, as mentioned above, could be a source of spurious correlation. As relevant, globally available, harmonized subnational data is sparse, this phenomenon could further investigated in regional validation studies, which we put for future research. Second, the assumption of the "gotta catch'em all" strategy of the dove constellation is overly simplistic and does not entirely reflect the actual capturing approach of Planet. However, following our analysis, Planet's capture strategy might focus on more rural, less developed areas. While this seems to be a possible positioning in the market, a more thorough market analysis could reveal more details on the dynamics in the satellite operator sector, which is beyond the scope of this paper. Third, a linear regression model does not properly capture the complex, non-linear relationships in the data. Figure~\ref{fig:linear_regression_plot} in the Appendix only partially supports this argument by displaying the actual vs. predicted plot for the fully-specified linear regression. Since the scatter plot shows a slight slope and heteroscedastic behaviour, thus hinting at non-linearities in the data, we repeat the analysis using a random forest model. As Figure \ref{fig:nonlinear_rf_regression_plot} in the Appendix shows that the random forest model is able to capture these non-linearities more accurately. The feature importance scores from the random forest model (cf. Figure \ref{fig:feature_importance_rf} in the Appendix) reflect effect sizes and significance levels from Table \ref{tab:regression_results_main}: a region's area size is the major determinant of historic image availability and both the population count and the SHDI play a statistically significant, but minor role in determining the historic image count, thus hinting at limited structural socio-economic biases in the historic image availability: less developed, more rural places have slightly less opportunities to reap the digital dividend of remote sensing due to limited data availability.

In addition, as unmodelled national-level effects such as policies on restricting the capture of high-resolution imagery, the presence of domestic satellite programs or varying success of sales efforts also potentially play a role in historic satellite image availability, we account for that by re-running our main regression with country-level fixed effects (see Table \ref{tab:total_regression_results_appendix_all_country_dummies} in the Appendix). By looking at the $R^2$ of the first regression, we observe that country-level fixed effects explain about 50\% of the sub-national variation in historic image availability in regression 1. However, when adding our main control variable -- \textit{Area Size} -- the additional explanatory power of the country-level fixed effects reduces substantially, implying that much of the fixed effects are related to the size and thus the size of the subnational areas of a country. Furthermore, replacing the composite index \textit{SHDI} with an indicator more related to economic power of a country -- the \textit{Income Index} published by GDL -- or running the main regression on very-high resolution imagery (i.e. $<= 0.5m$) only does not change the direction and significance of the effect (cf. with Tables \ref{tab:total_regression_results_appendix_all_incindex} and \ref{tab:total_regression_results_appendix_all_0-0.5mRes} in the Appendix). It can be noted though that the effect size more than doubles for the latter case, hinting at a stronger economic bias for very-high resolution data.

The Gini coefficient for the number of historic satellite images per $\text{km}^2$ across the 1726 subnational regions under study sorted by their respective SHDI is $0.64$ and therefore hints at large-scale inequalities in the distribution of available high-resolution satellite images: approx. 50\% of these images are captured in the 10\% of regions with the highest SHDI (cf. the Lorenz curve in Figure \ref{fig:lorenz_curve} in the Appendix). Conversely, the 40\% of regions with the lowest SHDI are represented on just 10\% of all available images. Interestingly, the Gini coefficient implies a stronger economic bias than the main regression shown in Table \ref{tab:regression_results_main} above. Reasons for this can be that the Gini coefficient reflects overall distributional inequality, not how well one variable linearly explains another. In other words, if the SHDI explains which regions tend to have more images but not how many exactly, the $\text{R}^2$ can still be comparatively low.

\subsection{Case studies on geopolitical conflict areas}\label{sec:results-cases}

As satellite imagery is also a major source of geospatial intelligence, we expect that geopolitical events such as armed conflicts also influence satellite image availability. To provide some initial visual evidence on this, we monitor satellite image availability over time for three recent conflict regions: Gaza, Sudan and Ukraine.

\subsubsection{Gaza war since October 2023}\label{sec:gaza}

\begin{figure*}[t]
    \centering
    \includegraphics[width=\textwidth]{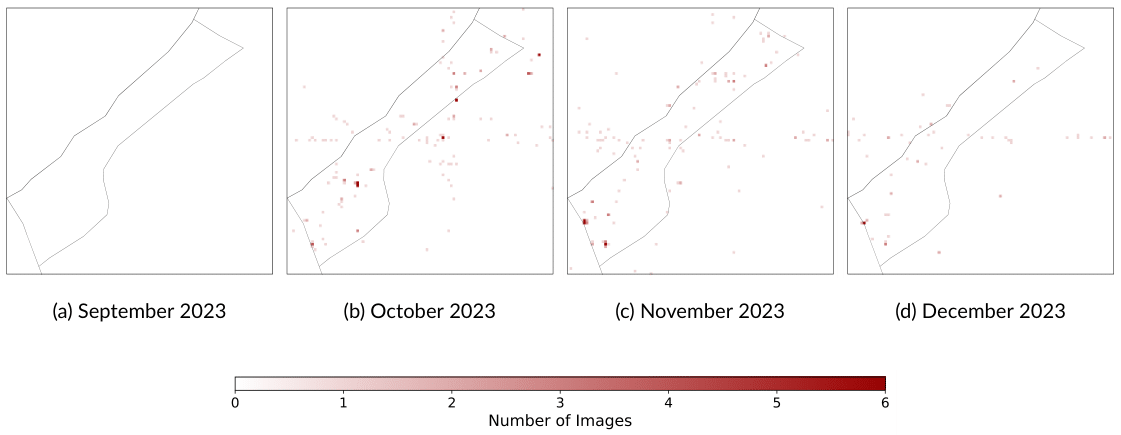}
    \caption{Satellite image availability for Gaza Strip from September 2023 to December 2023.}
    \label{fig:heatmap_gaza_grid}
\end{figure*}

Figure \ref{fig:heatmap_gaza_grid} shows how the availability of satellite imagery changes with the Hamas attack on Israel in October 2023.

While no images are available of Gaza for September 2023 and also images for prior months are sparse, image availability sharply increases in October 2023 - with hotspots at the site of the Re'im music festival massacre, near the kibbutz Nahal Oz and the cities Sderot and Khan Yunes. In November 2023, satellite image availability shifted towards Rafah and the southern border of Gaza, which reflects the shift in the battlelines at that time. In December 2023, overall satellite image availability reduced again.

\subsubsection{Civil war in Sudan since 2023}\label{sec:sudan}

\begin{figure*}[t]
    \centering
    \includegraphics[width=\textwidth]{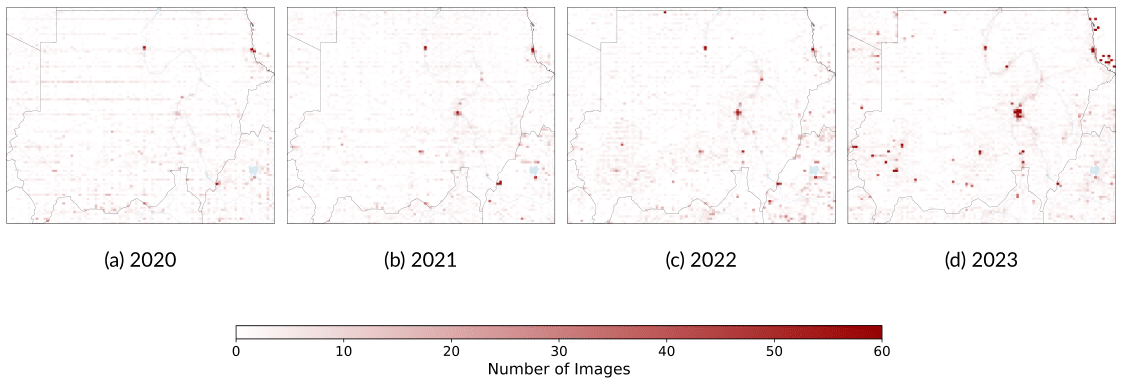}
    \caption{Satellite image availability for Sudan from 2020 to 2023.}
    \label{fig:heatmap_sudan_grid}
\end{figure*}

Before the outbreak of the recent Sudanese civil war in 2023, satellite image availability was low, with hotspots in Kerma, a major archaeological excavation site and Port Sudan, Sudan's main trade hub in 2020 and later on, in 2021 and 2022, in Khartoum, Sudan's capital, El-Obeid, a regional capital and the Grand Ethiopian Renaissance Dam near the Sudanese-Ethiopian border (which was being filled during that time) (see Figure \ref{fig:heatmap_sudan_grid}).

In 2023, satellite image availability increased strongly for Khartoum, which became the center stage of the conflict since. Additional hotspots interestingly developed in the waters around Port Sudan and in the paramilitary-held Darfur region.

\subsubsection{Russian invasion in Ukraine since February 2022}\label{sec:ukraine}

\begin{figure*}[t]
    \centering
    \includegraphics[width=\textwidth]{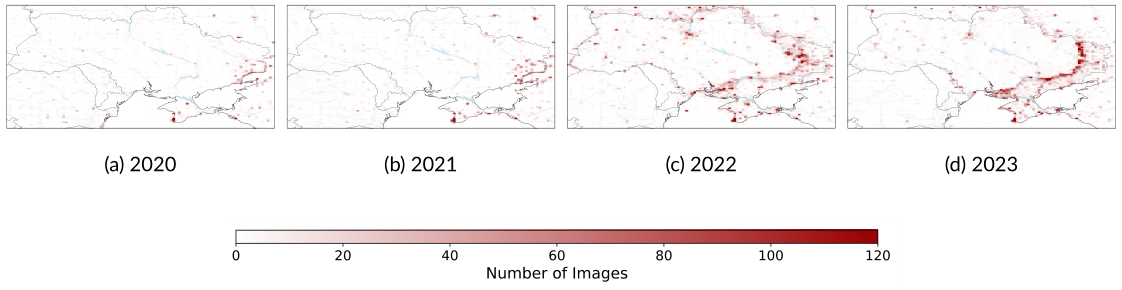}
    \caption{Satellite image availability for Ukraine from 2020 to 2023.}
    \label{fig:heatmap_ukraine_grid}
\end{figure*}

In Ukraine, as depicted in Figure \ref{fig:heatmap_ukraine_grid}, satellite image availability was considerably low in 2020, but rose at the end of 2021 at the border region of Ukraine and Russia and shifted simultaneously with the front line since.

From the visual patterns, it appears that the presence of armed conflicts is reflected in satellite image availability. Potential reasons for this are discussed in Section \ref{sec:discussion}.

\section{Discussion}\label{sec:discussion}

From Table \ref{tab:detailed_table}, we can see that available optical satellite imagery with a resolution below 10m remains behind what could potentially be captured. From a business perspective that is understandable as every capture incurs a cost (e.g. wear, downlink, online provision) and if expected profits from selling images are lower than their costs, private satellite operators may opt against offering these images in the first place. As the Gini coefficient and regression results in Table \ref{tab:regression_results_main} show, not only weather conditions such as varying cloud coverage apparently influence this trade-off, but also regional socio-economic situations. Socio-economic status, as crudely approximated by the SHDI in our study, significantly influences historic satellite image availability and, to a limited extent, explains the differences we observe between subnational regions, holding other things equal. This implies that poorer regions generally see less demand for satellite tasking, which in turn, due to the assumed business approach of most (very) high-resolution EO providers, leads to fewer archived images. This lack of demand can have multiple reasons, e.g. poorer regions have less infrastructure that is presumably worth monitoring from space such as ports or oil storage facilities and/or they lack of regional financial resources for satellite tasking, leading to downstream unmet latent demand. While the former may not necessarily hinder the ability to reap the digital dividend from earth observation for regional development, the latter presents a significant obstacle.


\section{Conclusion}\label{sec:conclusion}

In conclusion, we find that the availability of (very) high-resolution optical satellite imagery is not equally distributed across earth. While physical phenomena such as orbital trajectories can explain parts of that bias, differences are mainly due to business considerations based on the geo-political and socio-economic situation on the ground. While levels of the digital dividend a region can reap from leveraging EO data for local development is unknown, we expect that unmet latent demand leave poorer regions -- as measured by the SHDI in this study -- behind their economic growth potential. Even though this study offers an initial glimpse into these relationships, it is not without shortcomings: First of all, we cover the major, but not all providers of (very) high-resolution satellite imagery. Although we do not expect this to change the conclusions we can draw about the existence of coverage biases, we cannot rule it out. Second, as quantifying the economic value of using satellite imagery, what we call the digital dividend of EO data in this study, is beyond the scope of this paper, we cannot assess the significance of these biases for local development. Third and related to that, since we concentrate on (very) high-resolution imagery, we leave out the most prominent and used satellite programs, namely NASA's Landsat program and ESA's Sentinel program, out of the study. As government-led programs, they follow a "gotta catch 'em all" strategy, which significantly differs from the tasking-driven strategies of most privately-owned operators (with an exception of Planet). While many EO applications such as land use management, urban planning and disaster risk assessment, build on mid-resolution Landsat or Sentinel data and since these programs seem to be less affected by business model-driven coverage biases, the effect of these biases on the digital dividend might thus be less pronounced. Fourth, it is likely that our regression approach suffers from mis-specification such as omitted variable bias due to the scarcity of globally harmonized subnational data of relevance.

\section*{Funding}

Till Koebe and Ingmar Weber’s work is supported by funding from the Alexander von Humboldt Foundation and its founder, the Federal Ministry of Education and Research of Germany (Bundesministerium für Bildung und Forschung).


\section*{Appendix}\label{appendix}

\subsection*{Details on revisits and historic images}

\begin{table*}[ht]
\centering
\small
\begin{tabular}{lcccc}
\toprule
\textbf{Continent / Provider / Satellite} & \textbf{Avg. Daily Revisits} & \textbf{Avg. Daily Historic Images} & \textbf{Ratio} & \textbf{Resolution (m)} \\
\midrule
\textbf{Africa} & & & & \\
\quad \textbf{ImageSat} & & & & \\
\quad\quad EROS B & 364 & 5 & 0.014 & 0.8 \\
\quad \textbf{21AT} & & & & \\
\quad\quad OBJECT D & 308 & 23 & 0.075 & 0.5 \\
\quad\quad BEIJING 3B & 337 & 23 & 0.069 & 0.3 \\
\quad\quad TripleSat Constellation (DMC 3) & 1024 & 35 & 0.035 & 0.8 \\
\quad \textbf{Oneatlas} & & & & \\
\quad\quad Pléiades Neo & 673 & 156 & 0.233 & 0.3 \\
\quad\quad Pléiades & 679 & 116 & 0.172 & 0.5 \\
\quad\quad SPOT & 2418 & 67 & 0.028 & 1.5 \\
\quad \textbf{Planet Labs} & & & & \\
\quad\quad SkySat 1/2/C1-C13 & 4665 & 570 & 0.122 & 0.8 \\
\quad\quad SkySat C14-C19 & 2784 & 217 & 0.078 & 0.6 \\
\quad\quad FLOCK & 62202 & 564 & 0.010 & 3.9 \\
\quad \textbf{Maxar} & & & & \\
\quad\quad WorldView-1 & 347 & 77 & 0.222 & 0.6 \\
\quad\quad WorldView-2 & 351 & 77 & 0.221 & 0.5 \\
\quad\quad WorldView-3 & 339 & 62 & 0.184 & 0.4 \\
\quad\quad WorldView-4 & 344 & 27 & 0.080 & 0.3 \\
\quad\quad GeoEye-1 & 348 & 32 & 0.095 & 0.5 \\
\bottomrule
\end{tabular}
\caption{Daily revisits, historic images, and revisit-to-image ratio for Africa, by provider, and satellite.}
\label{tab:revisits_v_historic_africa}
\end{table*}

\begin{table*}[ht]
\centering
\small
\begin{tabular}{lcccc}
\toprule
\textbf{Continent / Provider / Satellite} & \textbf{Avg. Daily Revisits} & \textbf{Avg. Daily Historic Images} & \textbf{Ratio} & \textbf{Resolution (m)} \\
\midrule
\textbf{America} & & & & \\
\quad \textbf{ImageSat} & & & & \\
\quad\quad EROS B & 744 & 5 & 0.007 & 0.8 \\
\quad \textbf{21AT} & & & & \\
\quad\quad OBJECT D & 761 & 8 & 0.012 & 0.5 \\
\quad\quad BEIJING 3B & 770 & 12 & 0.016 & 0.3 \\
\quad\quad TripleSat Constellation (DMC 3) & 2283 & 46 & 0.020 & 0.8 \\
\quad \textbf{Oneatlas} & & & & \\
\quad\quad Pléiades Neo & 1523 & 246 & 0.162 & 0.3 \\
\quad\quad Pléiades & 1535 & 222 & 0.145 & 0.5 \\
\quad\quad SPOT & 5377 & 196 & 0.037 & 1.5 \\
\quad \textbf{Planet Labs} & & & & \\
\quad\quad SkySat 1/2/C1-C13 & 11427 & 1346 & 0.118 & 0.8 \\
\quad\quad SkySat C14-C19 & 4354 & 786 & 0.181 & 0.6 \\
\quad\quad FLOCK & 136568 & 1578 & 0.012 & 3.9 \\
\quad \textbf{Maxar} & & & & \\
\quad\quad WorldView-1 & 748 & 162 & 0.217 & 0.6 \\
\quad\quad WorldView-2 & 758 & 185 & 0.245 & 0.5 \\
\quad\quad WorldView-3 & 753 & 146 & 0.195 & 0.4 \\
\quad\quad WorldView-4 & 762 & 50 & 0.067 & 0.3 \\
\quad\quad GeoEye-1 & 773 & 63 & 0.082 & 0.5 \\
\bottomrule
\end{tabular}
\caption{Daily revisits, historic images, and revisit-to-image ratio for America, by provider, and satellite.}
\label{tab:revisits_v_historic_america}
\end{table*}

\begin{table*}[ht]
\centering
\small
\begin{tabular}{lcccc}
\toprule
\textbf{Continent / Provider / Satellite} & \textbf{Avg. Daily Revisits} & \textbf{Avg. Daily Historic Images} & \textbf{Ratio} & \textbf{Resolution (m)} \\
\midrule
\textbf{Asia/Pacific} & & & & \\
\quad \textbf{ImageSat} & & & & \\
\quad\quad EROS B & 503 & 9 & 0.018 & 0.8 \\
\quad \textbf{21AT} & & & & \\
\quad\quad OBJECT D & 508 & 97 & 0.192 & 0.5 \\
\quad\quad BEIJING 3B & 518 & 56 & 0.108 & 0.3 \\
\quad\quad TripleSat Constellation (DMC 3) & 1543 & 77 & 0.050 & 0.8 \\
\quad \textbf{Oneatlas} & & & & \\
\quad\quad Pléiades Neo & 1039 & 260 & 0.251 & 0.3 \\
\quad\quad Pléiades & 1033 & 271 & 0.262 & 0.5 \\
\quad\quad SPOT & 3598 & 210 & 0.059 & 1.5 \\
\quad \textbf{Planet Labs} & & & & \\
\quad\quad SkySat 1/2/C1-C13 & 7665 & 1123 & 0.146 & 0.8 \\
\quad\quad SkySat C14-C19 & 4884 & 769 & 0.158 & 0.6 \\
\quad\quad FLOCK & 90848 & 978 & 0.010 & 3.9 \\
\quad \textbf{Maxar} & & & & \\
\quad\quad WorldView-1 & 516 & 140 & 0.272 & 0.6 \\
\quad\quad WorldView-2 & 524 & 151 & 0.289 & 0.5 \\
\quad\quad WorldView-3 & 517 & 125 & 0.242 & 0.4 \\
\quad\quad WorldView-4 & 505 & 59 & 0.118 & 0.3 \\
\quad\quad GeoEye-1 & 507 & 61 & 0.121 & 0.5 \\
\bottomrule
\end{tabular}
\caption{Daily revisits, historic images, and revisit-to-image ratio for Asia/Pacific, by provider, and satellite.}
\label{tab:revisits_v_historic_ap}
\end{table*}

\begin{table*}[ht]
\centering
\small
\begin{tabular}{lcccc}
\toprule
\textbf{Continent / Provider / Satellite} & \textbf{Avg. Daily Revisits} & \textbf{Avg. Daily Historic Images} & \textbf{Ratio} & \textbf{Resolution (m)} \\
\midrule
\textbf{Europe} & & & & \\
\quad \textbf{ImageSat} & & & & \\
\quad\quad EROS B & 505 & 9 & 0.020 & 0.8 \\
\quad \textbf{21AT} & & & & \\
\quad\quad OBJECT D & 561 & 39 & 0.070 & 0.5 \\
\quad\quad BEIJING 3B & 514 & 26 & 0.052 & 0.3 \\
\quad\quad TripleSat Constellation (DMC 3) & 1589 & 29 & 0.019 & 0.8 \\
\quad \textbf{Oneatlas} & & & & \\
\quad\quad Pléiades Neo & 1063 & 95 & 0.089 & 0.3 \\
\quad\quad Pléiades & 1067 & 119 & 0.112 & 0.5 \\
\quad\quad SPOT & 3772 & 86 & 0.023 & 1.5 \\
\quad \textbf{Planet Labs} & & & & \\
\quad\quad SkySat 1/2/C1-C13 & 7992 & 487 & 0.061 & 0.8 \\
\quad\quad SkySat C14-C19 & 2350 & 309 & 0.132 & 0.6 \\
\quad\quad FLOCK & 93831 & 934 & 0.010 & 3.9 \\
\quad \textbf{Maxar} & & & & \\
\quad\quad WorldView-1 & 531 & 105 & 0.198 & 0.6 \\
\quad\quad WorldView-2 & 516 & 121 & 0.236 & 0.5 \\
\quad\quad WorldView-3 & 535 & 87 & 0.163 & 0.4 \\
\quad\quad WorldView-4 & 536 & 45 & 0.085 & 0.3 \\
\quad\quad GeoEye-1 & 533 & 35 & 0.067 & 0.5 \\
\bottomrule
\end{tabular}
\caption{Daily revisits, historic images, and revisit-to-image ratio for Europe, by provider, and satellite.}
\label{tab:revisits_v_historic_europe}
\end{table*}

\begin{table*}[ht]
\centering
\small
\begin{tabular}{lcccc}
\toprule
& \multicolumn{4}{c}{\textit{Dependent variable: Historic image count (Planet only)}} \\
\cmidrule(lr){2-5}
& (1) & (2) & (3) & (4) \\
\midrule
Constant & -0.009$^{**}$ & 0.001$^{}$ & -0.005$^{***}$ & -0.009$^{***}$ \\
& (0.004) & (0.005) & (0.002) & (0.002) \\
Subnational HDI & 0.025$^{***}$ & -0.029$^{***}$ & -0.004$^{}$ & -0.010$^{***}$ \\
& (0.006) & (0.009) & (0.003) & (0.003) \\
Latitude (abs.) & & 0.047$^{***}$ & 0.010$^{***}$ & 0.016$^{***}$ \\
& & (0.006) & (0.002) & (0.002) \\
Longitude (abs.) & & 0.036$^{***}$ & 0.006$^{***}$ & 0.006$^{***}$ \\
& & (0.005) & (0.002) & (0.002) \\
\# of households & & & -0.031$^{***}$ & -0.029$^{***}$ \\
& & & (0.005) & (0.005) \\
Area size & & & 0.868$^{***}$ & 0.869$^{***}$ \\
& & & (0.008) & (0.008) \\
Cloud coverage & & & & 0.013$^{***}$ \\
& & & & (0.002) \\
\midrule
Observations & 1726 & 1726 & 1726 & 1726 \\
$R^2$ & 0.009 & 0.059 & 0.875 & 0.878 \\
Adjusted $R^2$ & 0.009 & 0.057 & 0.875 & 0.878 \\
Residual Std. Error & 0.042 (df=1724) & 0.041 (df=1722) & 0.015 (df=1720) & 0.015 (df=1719) \\
F Statistic & 16.228$^{***}$ (df=1; 1724) & 35.942$^{***}$ (df=3; 1722) & 2405.248$^{***}$ (df=5; 1720) & 2068.890$^{***}$ (df=6; 1719) \\
\bottomrule
\textit{Note:} & \multicolumn{4}{r}{$^{*}$p $<$ 0.1; $^{**}$p $<$ 0.05; $^{***}$p $<$ 0.01} \\
\end{tabular}
\caption{OLS: Regressing the number of Planet image counts from the years 2017--23 on socio-economic and physical factors. All variables are normalized.}
\label{tab:total_regression_results_appendix_planet}
\end{table*}

\begin{table*}[ht]
\centering
\small
\begin{tabular}{lcccc}
\toprule
& \multicolumn{4}{c}{\textit{Dependent variable: Historic image count}} \\
\cmidrule(lr){2-5}
& (1) & (2) & (3) & (4) \\
\midrule
Constant & -0.010$^{**}$ & -0.001$^{}$ & -0.007$^{***}$ & -0.010$^{***}$ \\
& (0.005) & (0.005) & (0.001) & (0.002) \\
Subnational HDI & 0.031$^{***}$ & -0.026$^{***}$ & 0.001$^{}$ & -0.005$^{*}$ \\
& (0.006) & (0.009) & (0.003) & (0.003) \\
Latitude (abs.) & & 0.049$^{***}$ & 0.009$^{***}$ & 0.014$^{***}$ \\
& & (0.006) & (0.002) & (0.002) \\
Longitude (abs.) & & 0.042$^{***}$ & 0.008$^{***}$ & 0.008$^{***}$ \\
& & (0.005) & (0.001) & (0.001) \\
\# of households & & & -0.000$^{}$ & 0.002$^{}$ \\
& & & (0.004) & (0.004) \\
Area size & & & 0.916$^{***}$ & 0.917$^{***}$ \\
& & & (0.007) & (0.007) \\
Cloud coverage & & & & 0.011$^{***}$ \\
& & & & (0.002) \\
\midrule
Observations & 1726 & 1726 & 1726 & 1726 \\
$R^2$ & 0.013 & 0.069 & 0.914 & 0.916 \\
Adjusted $R^2$ & 0.013 & 0.067 & 0.913 & 0.916 \\
Residual Std. Error & 0.043 (df=1724) & 0.042 (df=1722) & 0.013 (df=1720) & 0.013 (df=1719) \\
F Statistic & 23.123$^{***}$ (df=1; 1724) & 42.433$^{***}$ (df=3; 1722) & 3641.115$^{***}$ (df=5; 1720) & 3131.862$^{***}$ (df=6; 1719) \\
\bottomrule
\textit{Note:} & \multicolumn{4}{r}{$^{*}$p $<$ 0.1; $^{**}$p $<$ 0.05; $^{***}$p $<$ 0.01} \\
\end{tabular}
\caption{OLS: Regressing the number of total image counts from the years 2017--23 on socio-economic and physical factors. All variables are normalized.}
\label{tab:total_regression_results_appendix_all}
\end{table*}

\begin{table*}[ht]
\centering
\small
\begin{tabular}{lcccc}
\toprule
& \multicolumn{4}{c}{\textit{Dependent variable: Historic image count}} \\
\cmidrule(lr){2-5}
& (1) & (2) & (3) & (4) \\
\midrule
Constant & 0.038$^{**}$ & -0.173$^{***}$ & -0.053$^{***}$ & -0.058$^{***}$ \\
& (0.016) & (0.019) & (0.008) & (0.008) \\
Subnational HDI & -0.034$^{*}$ & -0.036$^{*}$ & 0.013$^{*}$ & 0.011$^{}$ \\
& (0.021) & (0.019) & (0.008) & (0.008) \\
Latitude (abs.) & & 0.100$^{***}$ & 0.037$^{***}$ & 0.040$^{***}$ \\
& & (0.019) & (0.007) & (0.007) \\
Longitude (abs.) & & 0.428$^{***}$ & 0.101$^{***}$ & 0.101$^{***}$ \\
& & (0.027) & (0.011) & (0.011) \\
\# of households & & & 0.076$^{***}$ & 0.078$^{***}$ \\
& & & (0.006) & (0.006) \\
Area size & & & 0.928$^{***}$ & 0.931$^{***}$ \\
& & & (0.010) & (0.010) \\
Cloud coverage & & & & 0.018$^{***}$ \\
& & & & (0.004) \\
\midrule
Observations & 1726 & 1726 & 1726 & 1726 \\
$R^2$ & 0.542 & 0.620 & 0.939 & 0.939 \\
Adjusted $R^2$ & 0.489 & 0.576 & 0.931 & 0.932 \\
Residual Std. Error & 0.036 (df=1546) & 0.033 (df=1544) & 0.013 (df=1542) & 0.013 (df=1541) \\
F Statistic & 10.234$^{***}$ (df=179; 1546) & 13.928$^{***}$ (df=181; 1544) & 129.068$^{***}$ (df=183; 1542) & 129.975$^{***}$ (df=184; 1541) \\
\bottomrule
\multicolumn{2}{l}{\textit{Note: Country-level fixed effects included.}} & \multicolumn{3}{r}{$^{*}$p $<$ 0.1; $^{**}$p $<$ 0.05; $^{***}$p $<$ 0.01} \\
\end{tabular}
\caption{OLS with country-level fixed effects: Regressing the number of publicly available historic satellite images (without planet images) from the years 2017--23 on socio-economic and physical factors. All variables are normalized.}
\label{tab:total_regression_results_appendix_all_country_dummies}
\end{table*}

\begin{table*}[ht]
\centering
\small
\begin{tabular}{lcccc}
\toprule
& \multicolumn{4}{c}{\textit{Dependent variable: Historic image count}} \\
\cmidrule(lr){2-5}
& (1) & (2) & (3) & (4) \\
\midrule
Constant & -0.010$^{**}$ & -0.011$^{**}$ & -0.009$^{***}$ & -0.013$^{***}$ \\
& (0.005) & (0.005) & (0.002) & (0.002) \\
Income index & 0.042$^{***}$ & -0.006$^{}$ & 0.009$^{***}$ & 0.007$^{**}$ \\
& (0.007) & (0.009) & (0.003) & (0.003) \\
Latitude (abs.) & & 0.047$^{***}$ & 0.006$^{***}$ & 0.009$^{***}$ \\
& & (0.007) & (0.002) & (0.002) \\
Longitude (abs.) & & 0.054$^{***}$ & 0.014$^{***}$ & 0.013$^{***}$ \\
& & (0.005) & (0.002) & (0.002) \\
\# of households & & & 0.075$^{***}$ & 0.076$^{***}$ \\
& & & (0.005) & (0.005) \\
Area size & & & 1.034$^{***}$ & 1.035$^{***}$ \\
& & & (0.008) & (0.008) \\
Cloud coverage & & & & 0.008$^{***}$ \\
& & & & (0.002) \\
\midrule
Observations & 1726 & 1726 & 1726 & 1726 \\
$R^2$ & 0.023 & 0.085 & 0.913 & 0.914 \\
Adjusted $R^2$ & 0.022 & 0.083 & 0.913 & 0.914 \\
Residual Std. Error & 0.050 (df=1724) & 0.049 (df=1722) & 0.015 (df=1720) & 0.015 (df=1719) \\
F Statistic & 39.689$^{***}$ (df=1; 1724) & 53.104$^{***}$ (df=3; 1722) & 3626.481$^{***}$ (df=5; 1720) & 3054.818$^{***}$ (df=6; 1719) \\
\bottomrule
\textit{Note:} & \multicolumn{4}{r}{$^{*}$p $<$ 0.1; $^{**}$p $<$ 0.05; $^{***}$p $<$ 0.01} \\
\end{tabular}
\caption{OLS: Regressing the number of publicly available historic satellite images (without planet images) from the years 2017--23 on socio-economic and physical factors. All variables are normalized.}
\label{tab:total_regression_results_appendix_all_incindex}
\end{table*}

\begin{table*}[ht]
\centering
\small
\begin{tabular}{lcccc}
\toprule
& \multicolumn{4}{c}{\textit{Dependent variable: Historic image count ($0-0.5$ meter resolution)}} \\
\cmidrule(lr){2-5}
& (1) & (2) & (3) & (4) \\
\midrule
Constant & -0.015$^{**}$ & -0.007$^{}$ & -0.012$^{***}$ & -0.013$^{***}$ \\
& (0.006) & (0.006) & (0.002) & (0.002) \\
Subnational HDI & 0.055$^{***}$ & -0.012$^{}$ & 0.018$^{***}$ & 0.016$^{***}$ \\
& (0.008) & (0.012) & (0.004) & (0.004) \\
Latitude (abs.) & & 0.056$^{***}$ & 0.004$^{}$ & 0.006$^{*}$ \\
& & (0.008) & (0.003) & (0.003) \\
Longitude (abs.) & & 0.063$^{***}$ & 0.014$^{***}$ & 0.014$^{***}$ \\
& & (0.006) & (0.002) & (0.002) \\
\# of households & & & 0.103$^{***}$ & 0.104$^{***}$ \\
& & & (0.007) & (0.007) \\
Area size & & & 1.149$^{***}$ & 1.149$^{***}$ \\
& & & (0.011) & (0.011) \\
Cloud coverage & & & & 0.004$^{}$ \\
& & & & (0.002) \\
\midrule
Observations & 1726 & 1726 & 1726 & 1726 \\
$R^2$ & 0.024 & 0.085 & 0.885 & 0.885 \\
Adjusted $R^2$ & 0.023 & 0.083 & 0.885 & 0.885 \\
Residual Std. Error & 0.057 (df=1724) & 0.055 (df=1722) & 0.020 (df=1720) & 0.020 (df=1719) \\
F Statistic & 42.491$^{***}$ (df=1; 1724) & 53.273$^{***}$ (df=3; 1722) & 2648.491$^{***}$ (df=5; 1720) & 2209.375$^{***}$ (df=6; 1719) \\
\bottomrule
\textit{Note:} & \multicolumn{4}{r}{$^{*}$p $<$ 0.1; $^{**}$p $<$ 0.05; $^{***}$p $<$ 0.01} \\
\end{tabular}
\caption{OLS: Regressing the number of publicly available historic satellite images with a resolution $0-0.5$m from the years 2017--23 on socio-economic and physical factors. All variables are normalized.}
\label{tab:total_regression_results_appendix_all_0-0.5mRes}
\end{table*}

\subsection*{List of variables used in the linear regression}
\label{appendix:variables}

\begin{table*}[ht]
\centering
\begin{tabular}{lp{9cm}}
\toprule
\textbf{Variable} & \textbf{Description} \\
\midrule
\texttt{Historic image count} & Normalized count of satellite images for each region. Used as the dependent variable. \\
\texttt{Subnational HDI} & Normalized Subnational Human Development Index – a composite indicator of education, health, and standard of living. \\
\texttt{Latitude (abs.)} & Normalized absolute latitude of the centroid of the region. \\
\texttt{Longitude (abs.)} & Normalized absolute longitude of the centroid of the region. \\
\texttt{Area size} & Normalized area of the region (originally in square kilometers). \\
\texttt{\# of households} & Normalized number of households in the region. \\
\texttt{Cloud coverage} & Normalized average cloud coverage over the region. \\
\bottomrule
\end{tabular}
\caption{Variables included in the linear regression models. Normalizations are done using Min--Max scaling.}
\label{tab:regression_variables}
\end{table*}

\subsection*{Robustness of regression}
\label{appendix:robustness}

\begin{figure}[t]
    \centering
    \includegraphics[width=\linewidth]{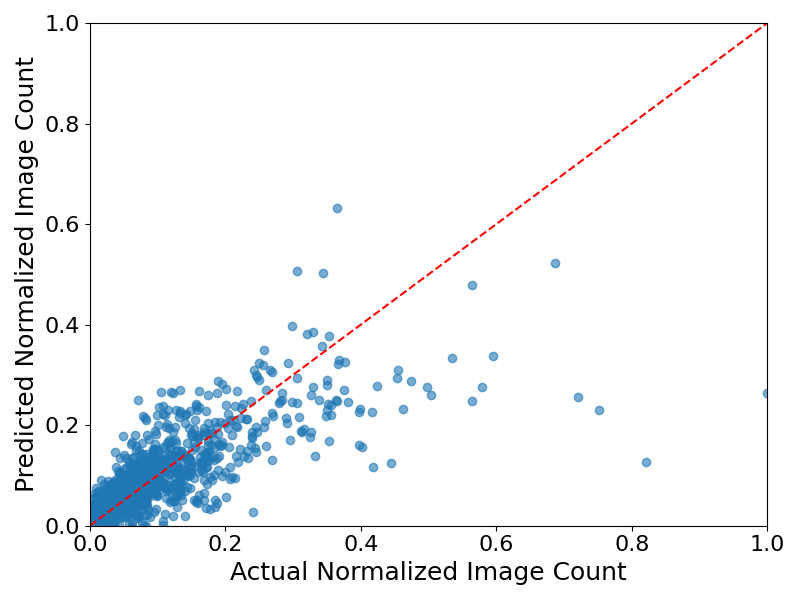}
    \caption{Actual vs. predicted normalized historic image counts from the linear regression model. The red dashed line represents perfect prediction.}
    \label{fig:linear_regression_plot}
\end{figure}

\begin{figure}[t]
    \centering
    \includegraphics[width=\linewidth]{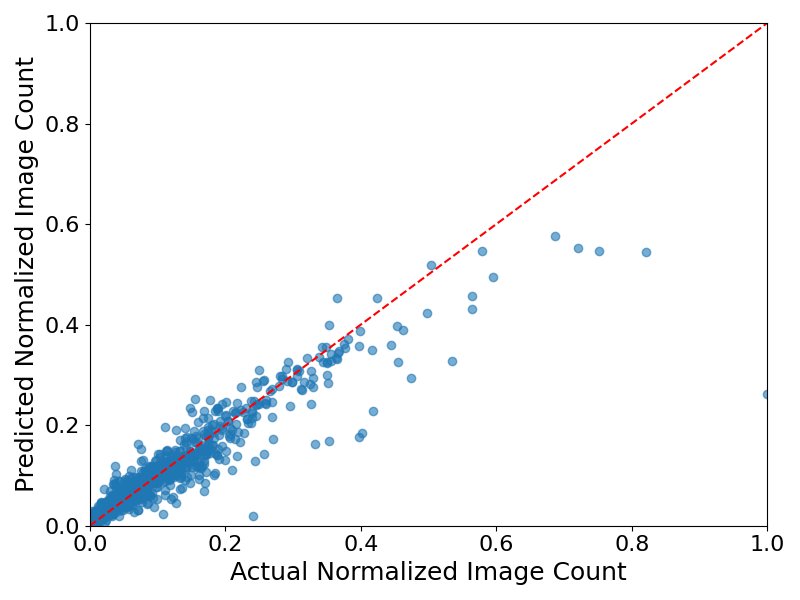}
    \caption{Actual vs. predicted normalized historic image counts from the Random Forest model. The red dashed line represents perfect prediction.}
    \label{fig:nonlinear_rf_regression_plot}
\end{figure}

\begin{figure}[ht]
    \centering
    \includegraphics[width=\linewidth]{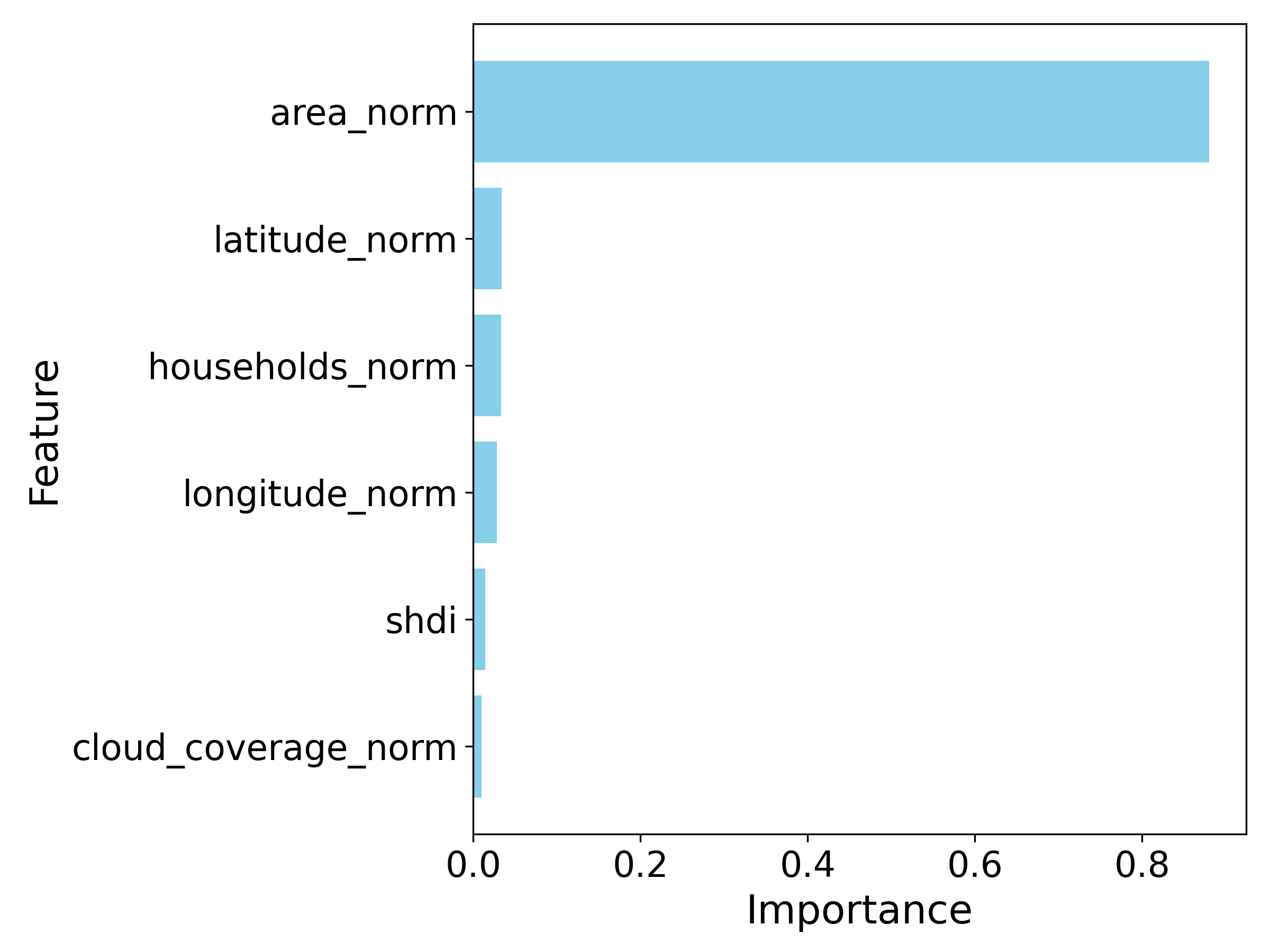}
    \caption{Feature importance scores for the Random Forest model predicting image counts.}
    \label{fig:feature_importance_rf}
\end{figure}


\begin{figure}[ht]
    \centering
    \includegraphics[width=0.5\textwidth]{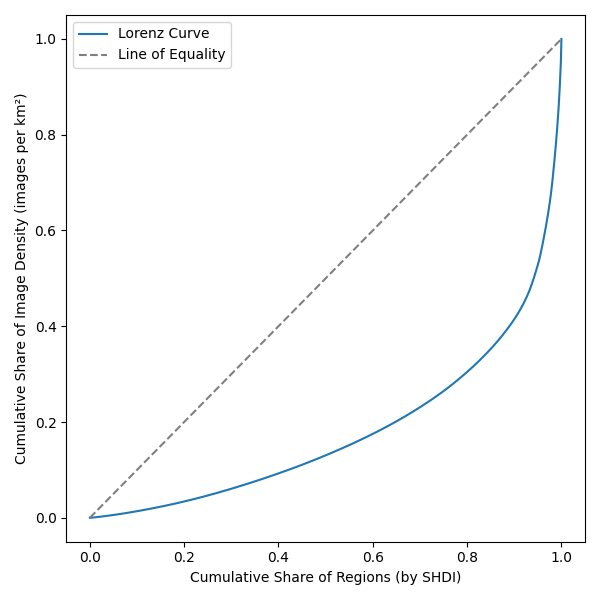}
    \caption{Lorenz curve showing the distribution of historic satellite image availability (images per km²) across regions, ordered by their Subnational Human Development Index (SHDI). The curve deviates notably from the dotted line of equality, indicating a higher concentration of high-resolution imagery per unit area in more developed regions. The Gini coefficient is 0.638.}
    \label{fig:lorenz_curve}
\end{figure}
\bibliography{aaai25}

\end{document}